# Recurrent network dynamics explain the time course of perceptual grouping in natural scenes

Sami Mollard[1*], Alekh K. Ashok[2], Lore Goetschalckx[2], Drew Linsley[2,3], Thomas Serre[2,3], Sander M. Bohte[4,5], Pieter R. Roelfsema[1,6,7,8*]

[1]Department of Vision & Cognition, Netherlands Institute for Neuroscience, Meibergdreef 47, 1105 BA Amsterdam, the Netherlands
[2]Department of Cognitive & Psychological Sciences, Brown University, Providence, RI 02912, USA
[3]Carney Institute for Brain Science, Brown University, Providence, RI 02912, USA
[4]Machine Learning Group, Centrum Wiskunde & Informatica, Science Park 123, 1098 XG Amsterdam, the Netherlands
[5]Swammerdam Institute for Life Sciences, University of Amsterdam, Science Park 904, 1090 GE Amsterdam, the Netherlands
[6]Laboratory of Visual Brain Therapy, Sorbonne Université, Institut National de la Santé et de la Recherche Médicale, Centre National de la Recherche Scientifique, Institut de la Vision, F-75012 Paris, France
[7]Department of Integrative Neurophysiology, Center for Neurogenomics and Cognitive Research, VU University, Amsterdam, the Netherlands
[8]Department of Neurosurgery, Academic University Medical Center, Amsterdam, the Netherlands

*Corresponding author

# Abstract

How the brain groups image elements into coherent objects in natural scenes remains unclear. Existing models for grouping in biological vision rely on simplified stimuli with explicit boundaries and cannot explain human behavior in naturalistic settings.

We propose a mechanistic framework in which recurrent interactions propagate enhanced neuronal activity within and between cortical areas. The framework links computational principles, neural circuitry and perceptual psychology. Local boundary signals govern early grouping, whereas later stages integrate top-down feedback carrying information about object identity, grouping features across internal edges into coherent object representations.

We instantiated this framework as a brain-inspired recurrent neural network trained to group features in natural images. The network learned to propagate enhanced activity across a cued object's representation, mirroring the brain's perceptual grouping mechanisms. The network's dynamics also predicted human reaction times. The framework links cortical dynamics to human vision and generates testable predictions for neurophysiological and psychophysical experiments.

# Introduction

Visual processing unfolds over successive processing phases. During an initial feedforward sweep, neurons in early visual cortex extract local features such as edges, textures and colors and transmit this information to higher visual areas that extract more complex feature constellations, including object shape and category. This first phase yields feature maps in lower areas and more abstract object-related signals in higher areas, but it does not yet specify how spatially distributed features are grouped into objects. This initial representation therefore differs from our perception of a world filled with spatially extended objects[1–6].

The brain must have powerful mechanisms to group features into coherent objects and segregate them from other objects and the background. Perceptual grouping unfolds over hundreds of milliseconds, substantially longer than the feedforward sweep[2], and likely depends on recurrent processing through feedback connections from higher to lower areas and horizontal connections within brain regions. These interactions bind spatially distributed features into coherent object representations.

Our understanding of the brain mechanisms for perceptual grouping could benefit from recent advances in artificial intelligence, which have produced highly accurate models that can delineate objects in complex natural scenes[7,8]. However, these models were not designed to explain the neural and attentional mechanisms that underlie perceptual grouping in humans, which remain only partially understood. Here, we use advances in image segmentation as a starting point for a mechanistic, biologically plausible model of perceptual grouping in naturalistic images.

During perceptual grouping, recurrent cortical interactions are thought to spread enhanced neuronal activity across an object's representation. Neuronal recordings in the primary visual cortex (area V1) have shown that response enhancement spreads automatically from attended elements to other elements linked by Gestalt cues, like collinearity, color similarity and common fate[9]. This activity labels the object's elements serially and corresponds, at the psychological level, to the spread of object-based attention[10–19]. The correspondence between object-based attention and the spread of enhanced neuronal activity has been demonstrated in the curve-tracing task, where participants trace a target curve while ignoring a distractor (Fig. 1A). Reaction times (RTs) scale approximately linearly with curve length, consistent with a serial process[12] in which attention propagates gradually along the curve[10]. In monkeys, the neuronal responses elicited by the target curve are enhanced relative to responses elicited by the distractor (Fig. 1B). The latency of the response enhancement increases with the distance from the start of the tracing process, in accordance with a serial labeling process. The latency also increases if the target and distractor curves are close together[15]. When the separation between curves narrows at a bottleneck (Fig. 1A), the propagation of the response enhancement is selectively delayed for neurons whose receptive fields (RFs) lie beyond the bottleneck.

The finding that the speed of propagation in the curve-tracing task depends on the distance between curves suggests that enhanced activity can propagate at multiple levels of the visual hierarchy[15]. When the target and distractor curves are close together, propagation may be restricted to lower visual areas such as the primary visual cortex (V1). Here, small RFs and horizontal connections between nearby visual field locations result in slow tracing. When the curves are farther apart, propagation can also occur in higher areas, whose neurons have larger RFs and whose horizontal connections span more distant locations, resulting in faster propagation. Consistent with this multi-scale account, RT patterns are well explained by a Growth-Cone (GC) model, in which propagation speed scales with the distance between the target and distractor curves (Fig. 1D).

Evidence for incremental grouping extends to two-dimensional shapes. A previous study[20] examined natural images containing two animals or vehicles and asked participants to report whether two cues fell on the same or different objects (the *same-different* task[12,20–22]). When the cues fell on the same object, RTs increased with the distance between them, again indicating an iterative grouping process. Notably, RTs also increased when images were presented upside down, a manipulation that impairs object recognition, suggesting that higher-level recognition processes facilitate perceptual grouping[23]. A subsequent study[21] compared variants of this task (Fig. 1E) and found that RT patterns were well explained by a variant of the GC model for line drawings of unrecognizable objects. This model constrained the GC size so that it remained within object outlines (Fig. 1F). However, the predictive power of the GC model[24–30] deteriorated in the presence of textures or internal edges, or when objects were readily recognizable (Fig. 1G), indicating a failure to account for grouping dynamics in more naturalistic settings. Furthermore, the GC models relied on explicitly defined object boundaries (e.g. line drawings), leaving open how grouping is achieved in natural images where objects have complex internal structure and their boundaries can be ambiguous.

Recent models have extended the study of perceptual grouping and object-based attention to natural images. Adeli et al.[22] modeled the spread of attention using similarities between image features extracted by a self-supervised transformer. Their model predicted human reaction times in a grouping task. Salehi et al.[31] used bidirectional recurrent gating to model attentional selection in complex scenes.

However, to our knowledge, no biologically grounded neural network has provided a mechanistic account of how enhanced activity propagates over object representations in natural images, where boundaries must be inferred and internal structure is complex.

Here, we investigate a candidate neuronal mechanism for perceptual grouping in naturalistic images. Using a recurrent neural network trained to group image elements, we address four questions: (1) how such grouping can be learned; (2) whether grouping emerges through the propagation of enhanced activity; (3) whether the resulting activity dynamics predict human RTs; and (4) how texture, internal edges and semantic information about object identity influence perceptual grouping.

We adapted GammaNet[32], a recurrent convolutional neural network that refines visual representations through horizontal and feedback connections, to perform a cued object-segmentation task based on the same–different paradigm. We trained it using contractor recurrent back-propagation (C-RBP)[33] , a variant of recurrent back-propagation (RBP) [34–36]. RBP avoids storing the complete activity trajectory required by back-propagation through time and permits synaptic updates based on locally available signals, as demonstrated by biologically motivated implementations[37]. The network learned to form perceptual groups in natural images by propagating enhanced activity and integrating local boundary cues with high-level semantic information. These results provide a computational framework linking image structure and semantic information to the propagation of enhanced neural activity and the resulting framework predicts perceptual grouping dynamics in human vision.

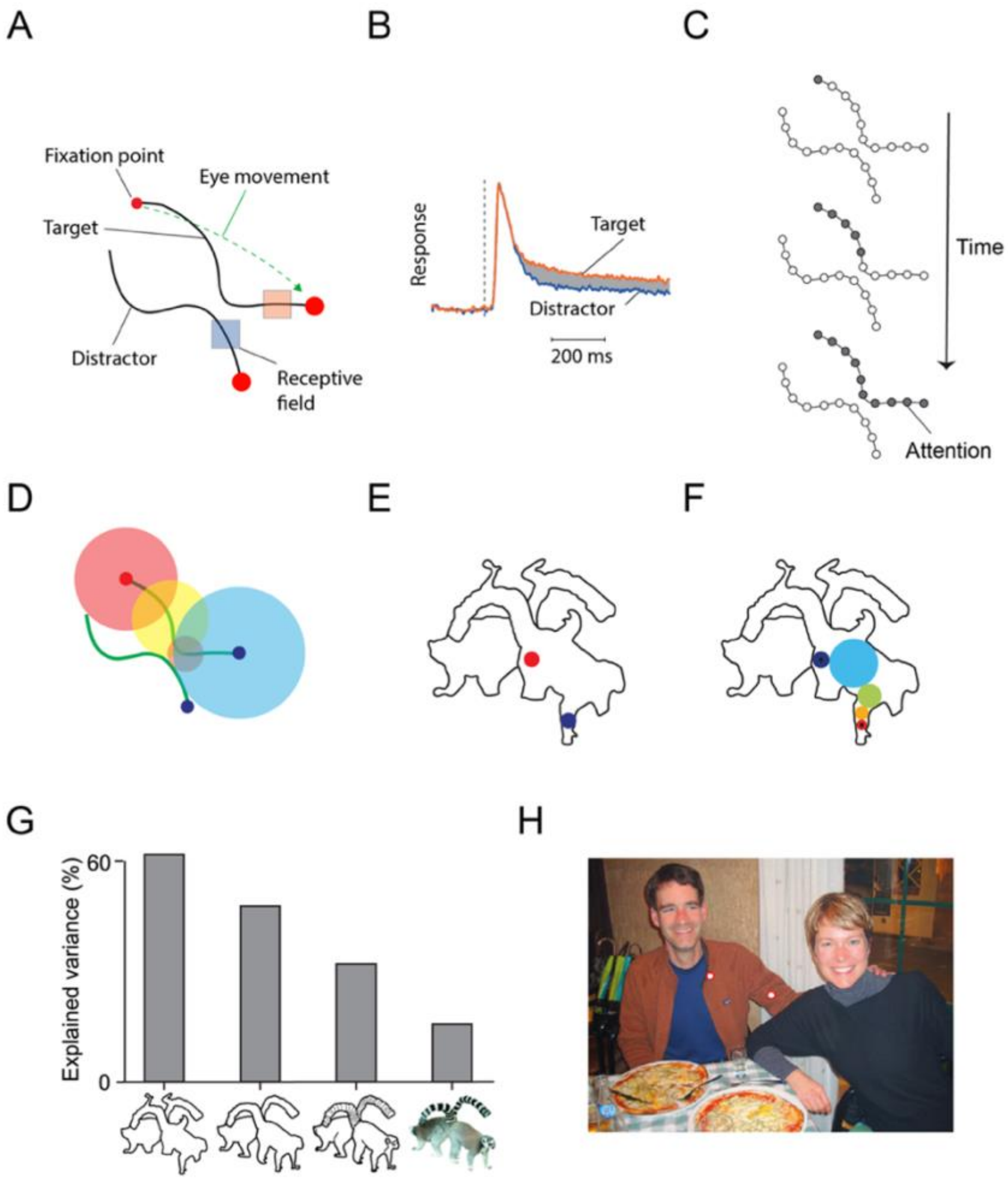

**Figure 1. Perceptual grouping unfolds incrementally over time. A.** Example curve-tracing stimulus. The subject fixates the red dot and makes an eye movement to the larger red circle connected to the fixation point by a target curve. Rectangles illustrate V1 receptive fields on the target (orange) and distractor curve (blue). **B**. Neuronal correlates of curve-tracing in the visual cortex. V1 neurons with RFs on the target curve (orange) have stronger responses than neurons with RFs on the distractor curve (blue) (adapted from ref.[38] ). **C.** Gradual spread of object-based attention. Enhanced neuronal activity propagates along the representation of the target curve (dark circles). **D.** Growth-Cone (GC) model of attentional propagation. Grouping speed is determined by the largest RFs that can be centered on the target curve without overlapping with the distractor curve. Grouping therefore accelerates when the separation between curves increases. **E.** Example stimulus from a same-different task probing perceptual grouping of two-dimensional objects. Participants report whether a cue (blue dot) falls on the same or a different object as the fixation point (red dot). **F.** The GC model predicts that grouping speed scales with object width, which determines the largest RFs that can participate in propagation. **G.** Performance of the GC-model across stimulus types. Its predictions are most accurate for unrecognizable line-drawings and deteriorate when objects are recognizable or contain internal contours, colours or textures (adapted from ref. [21]). **H.** Example stimulus from the perceptual grouping task used to measure human RTs[22]. Participants fixated the central cue and reported whether a second cue fell on the same object, here a person.

# Results

We asked whether a neural network model could learn to assign image elements to object instances, how its grouping signals evolve over time and whether its dynamics predicted human behavior. We trained a GammaNet[32] to group image elements using a semantically structured, retinotopic feature representation provided by a previously developed Feature Pyramid Network (FPN) [39] (Fig. 2B).

## Model and training procedure

The GammaNet received input from Panoptic R101-FPN[39], a network that was pretrained on the COCO dataset[40] to recognize and segment objects in natural scenes. The FPN extracts increasingly abstract feature representations and subsequently combines them with higher-resolution feature maps through upsampling and lateral connections. The output is a high-resolution, semantically structured representation that we used as input to GammaNet. Although the FPN is a feedforward network, its computational sequence resembles two phases of cortical processing: the initial feedforward progression through the visual hierarchy during which receptive-field sizes increase and category information emerges, followed by a feedback pass that integrates higher-level information into lower-level representations. Figure 2B illustrates this analogy using feedforward and feedback connections.

We will analyze this GammaNet input below and show that pixels with similar semantic roles, for example belonging to an animal's head or torso, cluster in feature space, and that object contours and internal boundaries can be decoded from this representation (Fig. 3E). The GammaNet must identify which pixels belong to the cued object. Class labels alone are insufficient when the image contains multiple objects of the same category.

The activities of the selected FPN layer were passed through a trainable projection layer that reduced the dimensionality and mapped it onto the GammaNet's input layer[30,32,41] (projection block in Fig. 2B). We trained the GammaNet to segment the cued object by assigning its category label to all pixels belonging to that object and labelling all other pixels as background (Fig. 2A).

The GammaNet consisted of four retinotopic layers at progressively coarser scales with horizontal gated recurrent units (hGRUs)[30] and connectivity inspired by the visual cortex. At each layer, the hGRUs form a retinotopic map with separate units for feedforward and feedback processing. In addition, horizontal connections enable long-range spatial interactions required for perceptual grouping[29,30,32,33,41]. Inhibitory interactions give rise to surround suppression, whereas excitatory interactions support the spread of activity across space, enhancing task-relevant representations[30,32] (see Methods and Fig. S1 for details). We evaluated the model's output by applying a linear readout to the activity in the lowest GammaNet layer, yielding pixel-wise category labels (Fig. 2B).

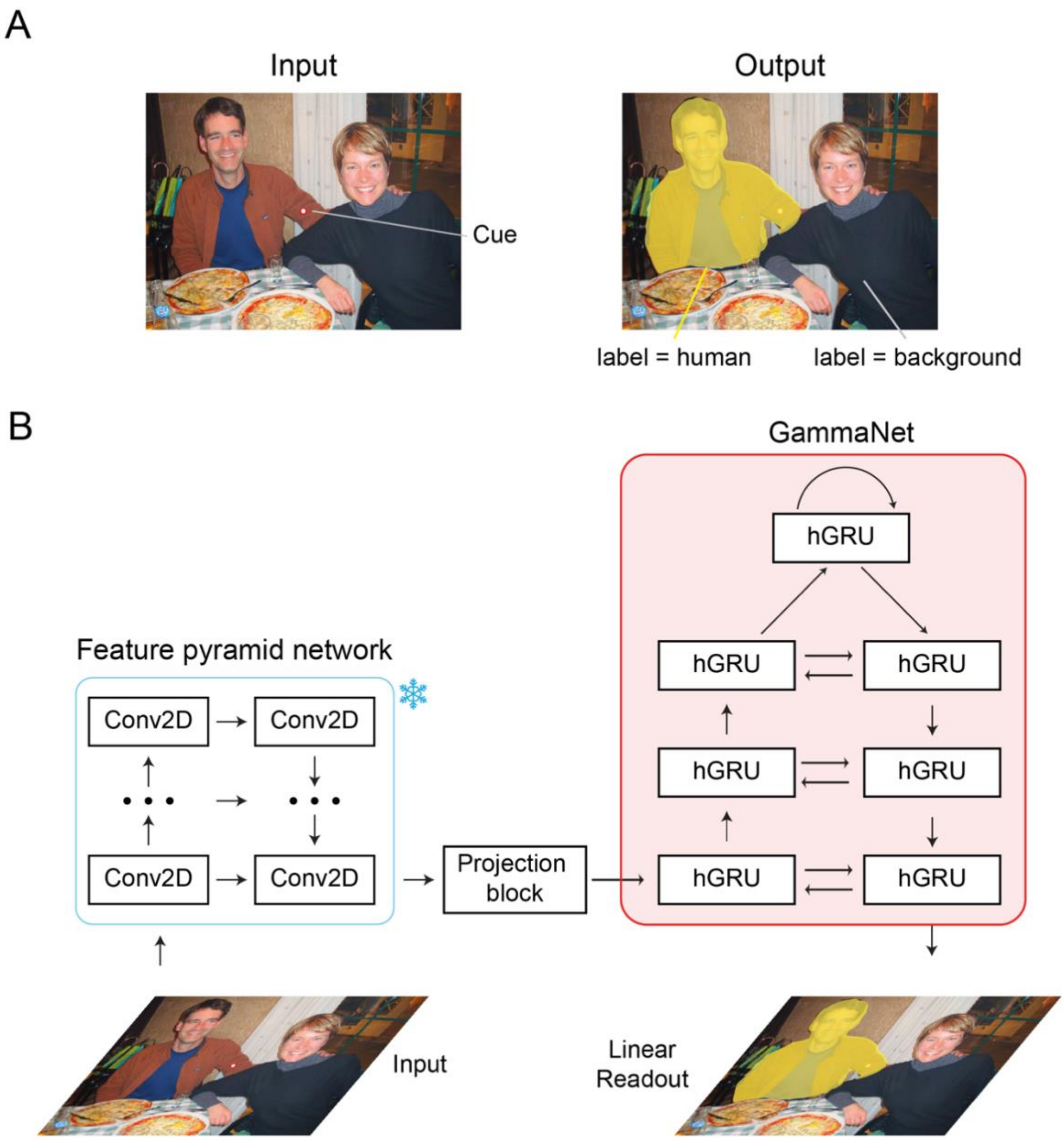

**Figure 2. The network and task used to examine perceptual grouping. A.** The model's task was to assign the correct class label (yellow) to all pixels of the cued object. **B.** Network architecture. We used the activity in an existing FPN specialized in semantic segmentation as input to the GammaNet via a projection block. This FPN is a purely feedforward network with skip connections, but its representations correspond to two cortical processing phases, in which feedforward signals first activate object class representations in higher areas and subsequently feed back to lower, retinotopically organized areas (upward and downward arrows). The ice crystal symbolizes that the FPN was "frozen"; we did not change its weights during training. The projection block reduces the dimensionality of the FPN output and provides high-resolution, semantically enriched information to the GammaNet. The GammaNet is a recurrent neural network with four hGRU layers with feedforward, horizontal and top-down connections. After recurrent processing by the GammaNet, the activity of the lowest hGRU layer is passed to a linear readout to generate per-pixel class predictions.

We used images from the COCO segmentation dataset[40] for which human reaction times have been measured in a same-different task[22]. In this task, one cue was placed on an object, and participants reported whether a second cue fell on the same or a different object (Fig. 1H)[20–22]. We presented the same images but used a segmentation objective. The GammaNet was trained to assign the cued object's class to every pixel belonging to that object, and to label all other pixels as background (Fig. 2A). Because the target mask specifies the relationship between the cue and every image location, each training example corresponds to supervision for many pairwise same-different judgments. Predicting the semantic class of the cued object, rather than a binary target label, further improved convergence.

We trained the GammaNet with contractor recurrent back-propagation (C-RBP)[33] , a variant of recurrent back-propagation (RBP)[34–36]. RBP computes gradients without storing the complete trajectory of recurrent activity, making the learning rule local in time. The RBP gradient can be evaluated by an accessory network that propagates credit-assignment signals, making the signals required for each weight update locally available at the synapse (see Methods)[37]. We trained five networks for 21,000 steps and report all five runs.

We assessed the models' segmentation performance using the intersection over union (IoU) between predicted and ground truth segmentation masks. IoU is the overlap between masks divided by their union, a standard metric in segmentation[42,43]. Across networks, the mean IoU was 0.74 ± 0.01 (mean ± SEM, N = 5 networks). For comparison, the widely used transformer-based "Segment Anything Model" (SAM) achieved an IoU of 0.81 ± 0.02 on the same images, and GammaNet's performance is only slightly lower. This performance gap is explainable because (1) the GammaNet has ~8 million parameters, whereas SAM has ~632 million parameters; (2) the GammaNet was exposed to ~5 million image-passes during training, including COCO pretraining of the FPN backbone plus fine-tuning, whereas the equivalent number is ~23 million for SAM; (3) the GammaNet was trained with the local C-RBP learning rule and SAM with regular back-propagation.

## Grouping emerges through the propagation of enhanced activity

We next examined how grouping unfolded over time by analyzing the pixelwise class labels read out from the lowest hGRU layer (Fig. 2). In an example image with two people (Fig. 3A), the GammaNets progressively labeled pixels of the cued person as human and those of the other person as background. Activity increased near the cue and spread across the cued person's representation (Fig. 3B). Enhanced activity propagated through all hGRU layers, at progressively coarser resolutions in higher layers (Fig. S3).

To quantify this propagation, we analyzed the images presented to human participants in ref. [22]. For each timestep, we computed the difference between the mean activity of lowest-layer hGRU units representing an object when it was cued and when another object was cued. This activity difference increased monotonically over time (Fig. 3C), indicating progressive enhancement of the cued object's representation relative to distractors. This temporal profile resembled activity dynamics in the visual cortex during perceptual grouping[15,44]. The training objective specified only the final semantic labels, not their temporal evolution. GammaNet's incremental propagation of enhanced activity therefore emerged during training on natural images.

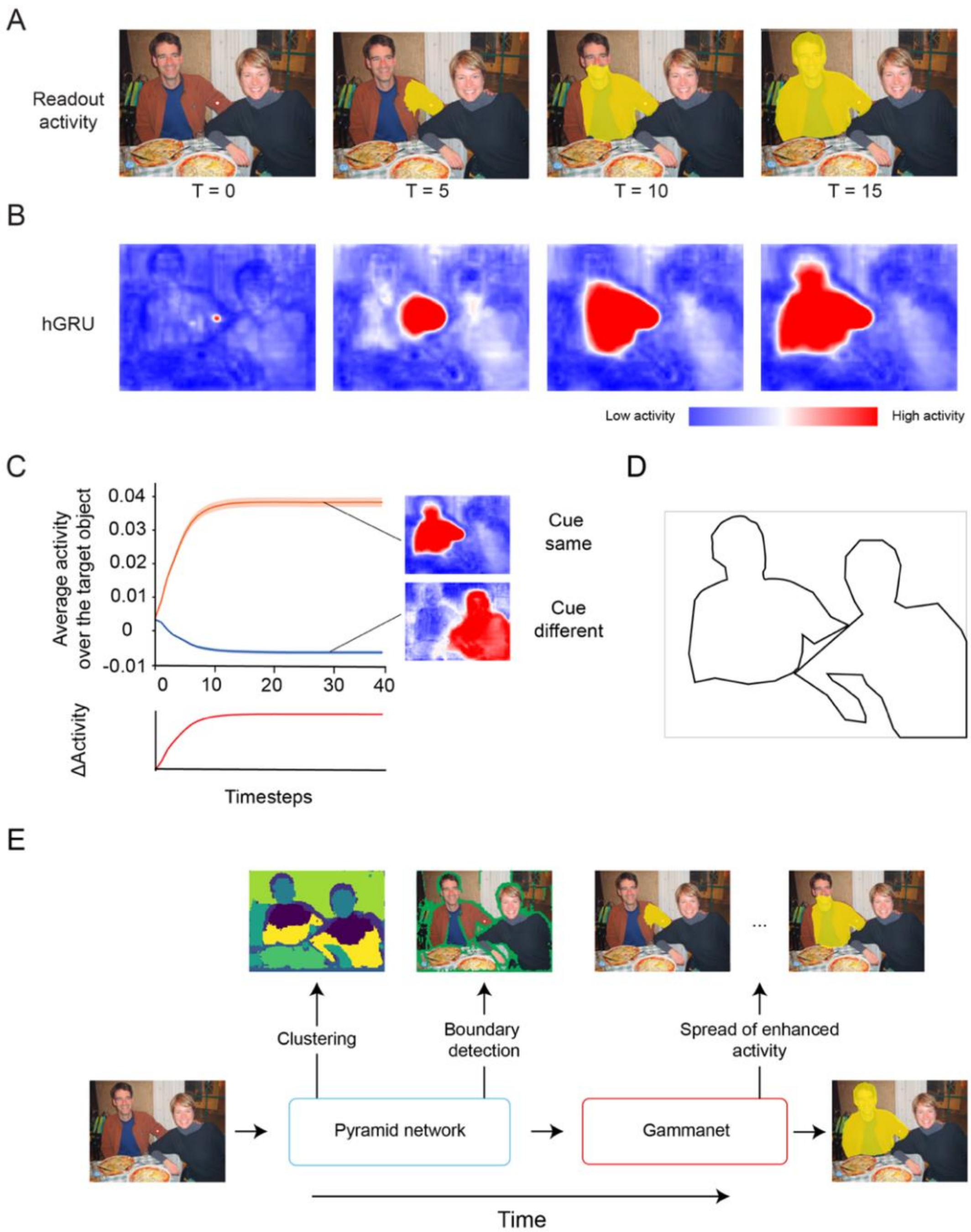

**Figure 3. Grouping emerges through the propagation of enhanced activity in the GammaNet. A.** Time-resolved pixelwise class predictions from the linear readout for an example image. Pixels belonging to the cued person are progressively assigned to the human class. **B.** Activity in the lowest hGRU layer for same image and timesteps. Enhanced activity (red) emerges near the cue and spreads across the representation of the cued object, while remaining low outside it. **C.** Population analysis across all test images. Mean activity of units with RFs on the target object (red) increases over time, whereas the activity of units representing a distractor object (blue) decreases. Lower panel, difference between these conditions. **D.** Contour-only version of the example image, used to evaluate models that require explicit object boundaries[28,29]. **E.** Summary of the processing stages. The FPN provides semantically structured features, in which pixels depicting similar object parts (e.g. torso, head) cluster in feature space and object boundaries can be decoded. The GammaNet propagates enhanced activity through the representation of the cued object, while external boundaries constrain its spread.

## Successive stages during grouping by the network

We analyzed the FPN's representation that was provided as input to the GammaNet (Fig. 2B). We started with K-means clustering of its feature maps (K=8 clusters), and observed clustering of pixels with similar roles, such as heads and torsos (Fig. 3E). However, this semantic information was implicitly encoded as vectors in a higher dimensional space.

We next examined whether the FPN also provided boundary information, using a linear readout that predicted object boundaries. We readily decoded external object contours and internal boundaries (Fig. 3E). Hence, the FPN input provides semantic and boundary information but it lacks an explicit representation of object-level grouping. The GammaNet used this input to propagate semantic class information from the cue to other regions occupied by the relevant object, linking spatially distributed elements into a unified representation[1–6].

## The GammaNet refines perceptual groups and crosses internal edges

Inspection of the grouping process showed that some edges that are initially represented as boundaries in the input to GammaNet are later overridden and incorporated into the cued object representation. These edges fall into two categories: interior edges caused by changes in illumination, color or texture within an object (Fig. 4A) and edges caused by occluding objects (Fig. 4A). Although the representation of interior edges initially resembles that of object borders, the GammaNet later relabels them as part of the object's interior. These edges create a conflict between the local evidence for a boundary and the global object-level semantic consistency. We therefore hypothesized that internal boundaries were overcome by semantic information in the GammaNet.

To examine how internal edges influence the grouping process, we removed selected internal edges in ten images with an image editing software (see Fig. 4B for an example image). The removal of internal edges only slightly improved the quality of the final segmentation (IoU 0.86 vs. 0.88; paired t-test, $t(4) = 4.47$, $p = 0.011$, $N = 5$ networks), and segmentation remained good in both conditions. The edges had, however, a much stronger influence on the grouping dynamics, which depended on their distance from the cue. When the cue fell near the edge between the person's legs in the example of Fig. 4B, the network initially treated the edge as an object boundary and propagated the grouping signal along it so that edge removal accelerated grouping (Fig. 4C). When the cue fell on the face, however, the same edge was crossed without delay (Fig. 4B). We quantified these effects by measuring arrival of the grouping signal at a target position on the far side of the edge (blue dots, Fig. 4B right). The internal edge in the example image of Fig 4C caused a mean delay of 4.0 ± 1.18 timesteps (mean ± SEM, N = 5 networks) when it was close to the cue, but only 0.6 ± 0.24 timesteps (mean ± SEM, N = 5 networks) when it was far from it (Fig. 4C).

We next tested whether this distance-dependent effect generalized across the ten images (Fig. 4D). There was an interaction between edge presence and cue-to-edge distance (repeated-measures ANOVA; $F(1,4)=7.9$, $p=0.048$). When the cue was near the edge, its presence delayed the spread of the grouping signal by an average of 2.3 ± 0.33 timesteps (mean ± SEM, N = 5 networks; $t(4)=6.8$, $p=0.0024$). When the cue was far from the edge, we detected no delay (0.02 ± 0.64 timesteps, mean ± SEM, N = 5 networks; $t(4)=0.03$, $p=0.98$)

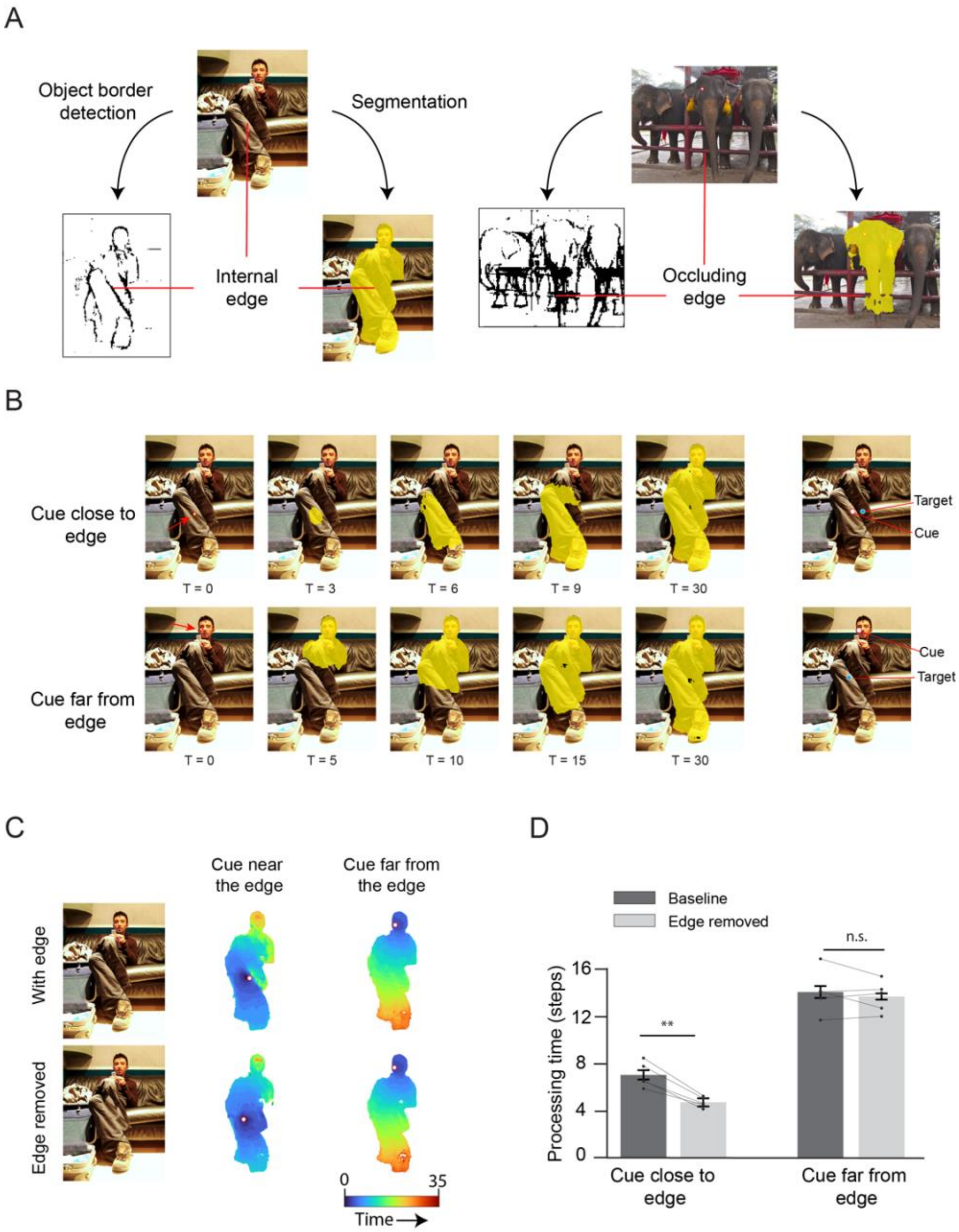

**Figure 4. Internal edges delay grouping when they are close to the cue. A.** Examples of ambiguous edges in natural images: an internal edge caused by contrast (left) and an edge caused by an occluding object (right). Both are detected as candidate object boundaries, but GammaNet ultimately groups the visible regions into the same object representation. **B.** Time course of grouping for cues placed near or far from an internal edge. **top,** When the cue is near the edge, the grouping signal initially propagates along the edge as though it were an object boundary before spreading across it. **bottom,** When the cue is far from the edge, the grouping signal crosses it without detectable delay. **right**, Blue dots indicate the target positions used to quantify propagation delays. **C.** Arrival-time maps for images with the selected edge present (top) and removed (bottom), and with the cue near (middle) or far from the edge (right). Colours indicate the timestep at which the grouping signal reached each pixel. **D.** Arrival times at the target positions for images with the edge present (dark grey) or removed (light grey), separately for cues near and far from the edge. Error bars, SEM. Dots represent individual models.

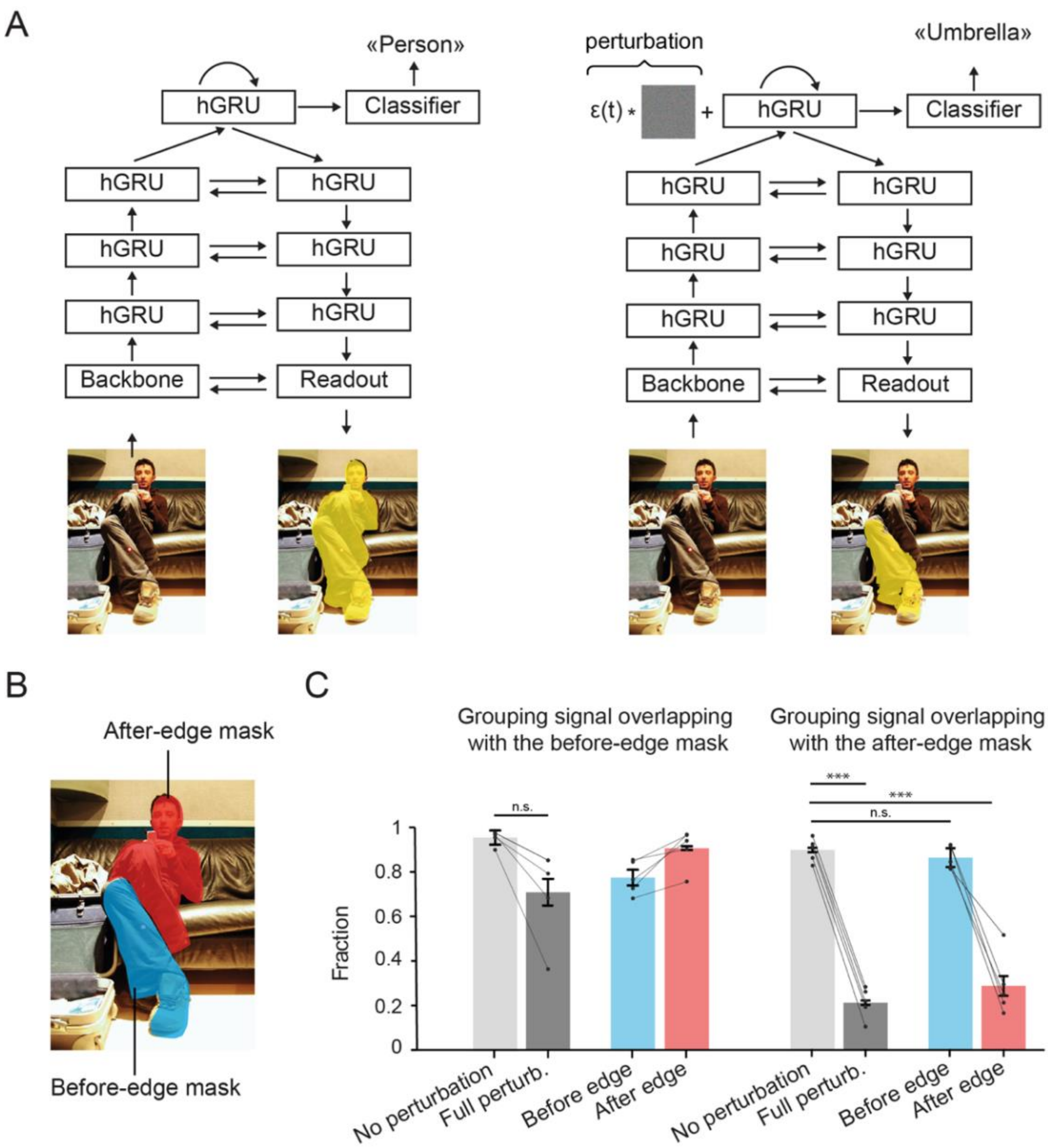

**Figure 5. Perturbing high-level category-related activity disrupts grouping across internal edges. A.** Schematic of the perturbation experiment. A linear classifier decoded the category of the cued object from activity in the highest hGRU layer. We perturbed activity in this layer so that the classifier misidentified the cued object without directly perturbing lower layers. In the example shown, the grouping signal reached the internal edge but failed to cross it. **B.** Definition of spatial masks. The *before-edge mask* comprised the region of the cued object on the same side of the internal edge as the cue; the *after-edge mask* comprised the region on the opposite side. **C.** Effect of spatially selective perturbations. Bars show the fraction of the before-edge mask (left four bars) and after-edge mask (right four bars) covered by the grouping signal. Perturbations were absent (light grey), applied to both regions (dark grey), restricted to the before-edge region (blue), or to the after-edge region (red). Dots represent individual models.

We hypothesized that category-related activity in the higher hGRU layers helps the GammaNet resolve internal edges. To test this hypothesis, we trained a linear classifier to predict the category of the cued object from activity in the highest layer. We then used the Basic Iterative Method (see Methods)[47] to perturb the activity in this layer until the classifier misidentified the person in our example image as “umbrella” (Fig. 5A). During perturbation, the grouping signal reached the internal edge but failed to cross it.

Across all test images, we applied the perturbation either to the region before the internal edge (before-edge mask), to the region beyond it (after-edge mask), or to both regions (full perturbation, Fig. 5B). Full perturbation strongly reduced coverage of the after-edge region (paired t-test, $t(4)=64$, $p<0.001$, t-tests were Bonferroni-corrected) (grey bars on the right of Fig. 5C). In contrast, the perturbation did not significantly reduce the grouping signal’s coverage of the before-edge region ($t(4)=3.23$, $p=0.22$) (grey bars on the left in Fig. 5C), suggesting that propagation up to the edge can rely on local feature information alone. The decrease in coverage of the before-edge region was significantly smaller than the decrease in coverage for the after-edge regions ($t(4)=6.35$, $p=0.022$).

We next examined the effects of the spatially selective perturbations. Restricting the perturbation to the after-edge region also reduced the coverage beyond the edge ($t(4)=14.67$, $p<0.001$), whereas perturbing the before-edge region produced no detectable reduction ($t(4)=0.97$, $p=1.0$). We also examined the effect of a transient perturbation of the top layer, which stalled the crossing of internal edges for the duration of the perturbation (Fig. S4). Hence, semantic information is necessary when internal boundaries need to be crossed. Furthermore, edge-crossing mostly relies on semantic information on the far side of the edge.

The perturbation results also provide insight into why internal edges delay grouping when the cue is nearby but have little effect when it is farther away (Fig. 4). Early propagation appears to depend primarily on local feature and boundary information. By the time the grouping signal reaches a more distant internal edge, activity representing a larger portion of the cued object may provide sufficient semantic evidence to cross it without delay. Together, these findings suggest that grouping within the GammaNet unfolds in stages (Fig. 3E). Early propagation follows local boundary evidence, whereas later category-related activity in higher layers enables the grouping signal to cross internal edges and establish a globally consistent object representation.

## GammaNet dynamics predict human reaction times during object parsing

We tested whether the GammaNet’s dynamics predicted human reaction times (RTs) in the same–different task of Adeli et al.[22]. Participants judged whether a second cue fell on the same object as the fixation point (Fig. 1H). We presented the same images but with only the fixation cue to the GammaNet, and estimated RTs as the time at which the segmentation signal reached the second cue, averaging across the five networks. The GammaNets generalized to held-out images, achieving 78% accuracy, compared to 90% in humans. These values are not directly comparable because the model had to satisfy a stricter segmentation criterion (Methods). The GammaNet explained 19.7% of the variance in human RTs, close to the estimated noise ceiling of 19.8% (Table 1). Thus, the GammaNet accounted for nearly all explainable variance, given the variability across participants. Because the GammaNet was trained on image segmentation rather than RT prediction, this correspondence emerged from the task, model architecture and learning constraints.

To examine how GammaNet's performance depends on its number of layers, which represent the image at distinct spatial scales, we also trained a version with 3 layers. This 3-layer GammaNet explained significantly less variance than the 4-scale model ($p < 10^{-6}$).

We next compared the GammaNet predictions with previous models of incremental grouping. Adeli et al.[22] modeled human RTs during grouping in natural images using affinity between image regions, which is a measure of feature similarity between representations, in pretrained transformers and ResNets. An iterative selection procedure, initialized at the cue location, successively added image patches based on affinity. The temporal grouping dynamics were therefore generated by this procedure rather than by the dynamics of the neural network itself. The model accounted for 7.3% of the variance in human RTs, which is less than the GammaNet ($p=2.8 \cdot 10^{-4}$) (Table 1). These researchers recently tested additional transformer and convolutional networks adding a readout network trained to predict human RTs from the network features[48]. Their best model now explained 20.5% of the variance in the human RT data, comparable to the GammaNet ($p=0.82$). In contrast, the GammaNet was never trained on RT data. It was trained only to segment images, and it directly modeled the propagation of the grouping signal through its recurrent dynamics. Its correspondence with human RTs therefore emerged entirely from learning to segment images.

Other previous models[15,21,28,29] only processed contour-only stimuli with explicit object boundaries (Fig. 3D) and could not overcome internal edges. One of these models is the Growth-Cone (GC) model[15,21], which provides a descriptive account of attentional propagation (Table 1). Mollard et al.[28] implemented the GC model as a neural network that learned to group image elements using reinforcement learning, whereas Goetschalckx et al.[29] trained an hGRU model to group image regions in line drawings. To compare these models with the GammaNet, we generated contour-only versions of the natural images by removing colors, textures and internal image structure (Fig. 3D). For these simplified stimuli, the GC and the Mollard et al.[28] models accounted for similar amounts of variance as the GammaNet’s did for the original images, whereas the model of Goetschalckx et al.[29] performed worse (Table 1). Thus, GammaNet extends previous accounts of incremental grouping to natural images, predicting human grouping times without requiring predefined object outlines or training on human RTs.

| Model | Remarks | Explained variance (95% CI) | p-value |
|---|---|---|---|
| Noise limit | | 19.8% (18.2-21.4%) | 0.99 |
| GammaNet, 4 layers | The present model | 19.7% (12-27%) | (reference) |
| GammaNet, 3 layers | | 4.2% (0-8.3%) | $8.4 \cdot 10^{-7}$ |
| Growth-Cone model | Hand engineered model operating on object boundaries | 16.8% (9.4-24%) | 0.45 |
| Mollard et al. (2026); ref. 28 | Neural implementation of the Growth-cone model. | 15.7% (8.5-23%) | 0.29 |
| Goetschalckx et al. (2023); ref. 29 | Recurrent network trained on object boundaries | 6.1% (1.1 – 11.1%) | $3.9 \cdot 10^{-5}$ |
| Adeli et al. (2023); ref. 22 | Affinity analysis in pretrained DNNs | 7.3% (2-12%) | $2.8 \cdot 10^{-4}$ |
| Adeli et al. (2026); ref. 48 | Model trained to fit human RTs directly | 20.5% (13-28%) | 0.82 |

**Table 1. Variance in human RTs explained by grouping models.** Human RT data are from Adeli et al.[22]. Among the models presented here, the GammaNet is the only neural network that groups regions directly in natural images. Adeli et al.[22] derived processing times from affinities between image regions in pretrained DNN representations and Adeli et al. [48] trained a read-out network on self-supervised transformer features to predict human RTs directly; the remaining models operated on contour-only stimuli (Fig. 3D). Values are explained variance with 95% confidence intervals. P-values are from Z-tests comparing each estimate with the four-scale GammaNet, which served as the reference model. For each model with several trained instances (all except the Growth-Cone model), the best instance is shown; all models were evaluated on the same trials.

## Discussion

We developed a recurrent neural network that learns to progressively group spatially distributed image elements into objects within natural images . The GammaNet receives input from a FPN and uses hGRUs[29,30,32,33,41] to implement horizontal and top-down interactions. The FPN input is high-dimensional, but contains decodable semantic and boundary information: related image regions have similar representations but are not yet explicitly grouped into object-level representations. Because the input can contain multiple objects of the same category (e.g. Fig. 3A), the FPN input does not specify which image regions are part of the same object.

The cue locally increases activity in the GammaNet, which then spreads through recurrent interactions. Early propagation follows local feature and boundary evidence, so that internal edges can temporarily halt the grouping signal and segregate regions belonging to the same object. Later, category-related activity in higher hGRU layers enables propagation across these edges. The GammaNet thereby combines local boundary evidence with high-level semantic information to establish coherent object representations.

The model's processing stages resemble proposed stages of perceptual grouping in the visual cortex[17,44,49]. During an initial feedforward processing phase, early visual areas encode elementary features such as edges[46] and higher areas encode progressively more complex shape information[1,50]. V1 neurons respond to abrupt feature changes[51], including true object boundaries[46] and internal edges, such as the edge between jacket and T-shirt in Fig. 3A. These early responses partition the scene into proto-objects[45] whose constituent regions have not yet been fully integrated.

Recurrent processing can subsequently resolve the ambiguities. Early V1 border responses are followed by a later phase in which responses to true object boundaries become stronger than responses to non-object borders[46]. Feedback also relays scene-related and semantic information to early visual cortex, including retinotopic regions without direct visual stimulation[52–54]. During this later, recurrent response phase, enhanced activity gradually spreads across the cortical representations of image elements belonging to the same object[1,9,15,44,55]. Consistent with these neuronal processes, psychophysical studies show that object-based attention spreads incrementally across an object, with longer processing times when task-relevant image elements are farther apart[10,21].

Most earlier computational models of incremental grouping were developed for simplified stimuli, such as isolated curves or object outlines on uniform backgrounds (Fig. 3D)[15,17,21,24,25,27,29,56]. The GC model, for example, spreads grouping signals across multiple spatial scales and accounts well for human RTs in grouping tasks[11,15,21]. These earlier models treat object boundaries as given, and do not address the distinction between object boundaries and internal edges. Instead, the GammaNet combines local boundary evidence with semantic information, enabling grouping to cross internal edges while respecting the object boundaries. It thereby extends the previous models of incremental grouping to natural images.

Our results also extend the findings by Adeli et al.[22,48] who predicted human RTs based on feature similarities between image regions in natural images using a transformer architecture. They used an iterative procedure or an additional neural network applied to the static transformer representations. Another promising approach[31] used a recurrent model to model attentional selection, but did not consider the time-resolved dynamics of perceptual grouping in natural scenes. The GammaNet goes beyond these previous studies by modeling the dynamics of perceptual grouping, propagating enhanced activity across the representation of the relevant object.

Modern deep neural networks, including Mask2Former[7] and SAM[8] , achieve high accuracy in natural image segmentation. These systems are optimized for benchmark segmentation performance but were not developed to explain grouping mechanisms in the visual cortex or the patterns of human RTs. Perceptual grouping in biological vision of spatially extended stimuli occurs for only one object at a time[57]. Accordingly, the GammaNet also spreads the grouping signal for only one object at a time, and accounts for human RTs, unlike the deep networks specifically optimized for image segmentation.

The GammaNet was trained with C-RBP[33], a contractive variant of RBP[34–36]. RBP is local in time because learning depends on equilibrium activity rather than on error signals propagating through previous network states. Its weight updates are also local in space once the postsynaptic credit signal is available. This contrasts with back-propagation through time[27,33], which is therefore unlikely to describe learning in the brain[58,59]. Importantly, Brosch et al.[37] showed how an accessory network can generate the required synapse-specific credit signals in a reinforcement-learning setting. The resulting RELEARNN rule combines these local credit signals with a global reward-prediction error, enabling a recurrent network to learn contour grouping and the gradual spread of enhanced activity. RBP-type credit assignment therefore provides a biologically motivated route by which the grouping dynamics expressed by GammaNet could be learned.

The present model has at least two limitations. First, GammaNet receives a semantically structured, high-resolution representation from a separately pretrained and frozen FPN. Future studies could examine whether this input stage and the recurrent grouping dynamics can be learned jointly in an end-to-end trainable architecture. Second, the GammaNet does not explicitly model border ownership signals, which indicate which side of a contour belongs to an object[60,61]. Although the GammaNet implicitly assigns boundaries to the selected object, future extensions could model the emergence of border-ownership signals and investigate how they interact with recurrent grouping.

The model makes experimentally testable predictions. Firstly, an ambiguous internal edge should delay perceptual grouping when it is close to the cue, but produce little or no delay if it is farther away. Grouping may reach a nearby internal edge before semantic feedback information is available in lower visual areas, whereas this feedback can already be available by the time it reaches a more distant edge. Second, information about the category of the cued object should become decodable from early visual cortex, including area V1. Category information should first emerge near the cue and then spread across the object's retinotopic representation.

More broadly, our results show how perceptual grouping in natural images can be understood as a dynamic process in which activity propagates under the influence of top-down semantic signals. This framework offers a promising step toward bridging the gap between biological vision and artificial systems, increasing our understanding of how the brain could construct object representations in natural scenes.

## Acknowledgments

This work was supported by a grant from the European Research Council [grant 101052963 NUMEROUS], the European Union Horizon 2020 Framework Program under specific grant agreement 945539 "Human Brain Project", an NWO crossover [grant 17619 'INTENSE'] and DBI2, a Gravitation program of the Dutch Ministry of Science, ONR (N00014-24-1-2026), NSF (IIS-2402875) and the ANR-3IA Artificial and Natural Intelligence Toulouse Institute (ANR-19-PI3A-0004).

# Methods

## Network Architecture

We generated training images from the COCO segmentation dataset[40] by adding cues at random locations on objects, following the paradigm of ref. [22]. The input to the GammaNet was provided by a Panoptic R101-FPN network, a feedforward FPN with skip connections trained on a panoptic segmentation task[39], which unifies semantic segmentation (assigning class labels to pixels) and instance segmentation (separating objects of the same class)[62] (Fig. 2B). We obtained the model weights from the Detectron2 model zoo and did not modify them during training of the GammaNet.

The FPN processes the input image through a forward pathway that extracts features and progressively encodes object class information at coarser spatial scales. This is followed by an upsampling pathway, in which semantically rich representations are projected to higher-resolution feature maps by combining information across scales[63].

We used the highest-resolution feature map (P2) as input to the GammaNet. This representation preserves fine spatial detail while incorporating semantic information, making it suitable for pixel-level perceptual grouping. We chose to only provide this input rather than all FPN feature maps to the corresponding GammaNet layers to prevent mixing of static FPN representations with the evolving GammaNet state. In our model, the GammaNet's dynamics evolve independently once initialized with the FPN output. The P2 feature map, of size $H/4 \times W/4$ with 128 channels, was passed through a $1\times1$ convolutional layer to reduce the dimensionality to 64 channels, followed by a ReLU nonlinearity.

The GammaNet consisted of four layers of stacked hGRU units organized across four spatial scales, with downsampling and upsampling operations enabling information exchange across resolutions. We explored variants with two to five hGRU layers and found that four layers yielded the best performance. Architectural details and hyperparameters are summarized in Table S1.

The model was trained to perform a multi-class segmentation task in which pixels belonging to the cued object were assigned their class label, while all other pixels were assigned to the background class. In terms of task structure, this formulation is equivalent to solving the same–different task for all possible pairs of cue locations: for any given pixel, the model effectively predicts whether it belongs to the same object as the cue or not. We note, however, that the segmentation objective provides a richer supervisory signal than a binary same/different task, since correct predictions must also carry the appropriate class label. We used a fixed set of 13 object classes determined by the selected image set (see below and Fig. S2) plus a background class. We note that it is straightforward to include additional classes.

The activity of the lowest hGRU layer gave rise to per-pixel class probabilities via a readout with two $1\times1$ convolutional layers followed by bilinear upsampling. The readout specified the probability of the 14 classes at each pixel as a 14-dimensional vector. The network was run for 40 timesteps although most stimuli were segmented in fewer timesteps. After the final timestep, a pixel was classified as 'same' if the highest probability class matched the class of the cued object, and as 'different' otherwise.

During training, the target class of pixels of the cued object was its class identity and the target class of all other pixels was 'background'. Training was performed using a per-pixel cross-entropy between predicted and ground truth class probabilities:

$$\mathcal{L} = -\sum_{i=1}^{H}\sum_{j=1}^{W}\sum_{k=0}^{K} y_{i,j,k} log\hat{y}_{i,j,k} \tag{1}$$

where *H* and *W* denote image height and width, *K* the number of classes (14), $y_{i,j,k} \in \{0,1\}$ is the one-hot ground truth label and $\hat{y}_{i,j,k} \in [0,1]$ the predicted class probability.

We quantified segmentation performance using the intersection-over-union (IoU) between predicted and ground-truth segmentation masks, where an IoU of 0 indicates no overlap and an IoU of 1 indicates perfect overlap.

## hGRUs

The dynamics of the hGRU units[29,30,33] are defined by a recurrent update of their hidden state over time. At each timestep $t$, the hidden state $H[t]$ is updated based on the previous state $H[t-1]$ and an external input $Z$, allowing the unit to integrate information across space and time. Each update consists of two stages, a suppressive stage followed by a facilitative stage, which together regulate the propagation of activity through the network. We used Group Normalization (GN) throughout to stabilize training and mitigate exploding or vanishing gradients. The full update process is illustrated schematically in Fig. S1.

*Suppression stage*: At every timestep, the hGRU first receives suppressive influences from the previous state. These influences are controlled by a suppression gate $G^S$, which determines how strongly the previous state influences subsequent computations:

$$G^S = sigmoid(U^S * H[t-1]), \tag{2}$$

where $H[t-1] \in \mathbb{R}^{H \times W \times C}$ denotes the hidden state at the previous time step, $*$ denotes convolution, $U^S \in \mathbb{R}^{1 \times 1 \times C \times C}$ is a set of learnable convolutional kernels and *C* the number of channels.

The gated hidden state was then transformed through a convolutional suppression pathway:

$$C^S = GN(W^S * H[t-1] \odot G^S), \tag{3}$$

where $W^S \in \mathbb{R}^{K \times K \times C \times C}$ denotes learnable convolutional kernels, $\odot$ denotes element-wise multiplication and *GN*(·) group normalization, normalizing across groups of channels to improve the stability of learning. We used spatial kernels with *K*=5 for horizontal interactions and *K*=1 for feedback, channel mixing interactions.

The external input $Z \in \mathbb{R}^{H \times W \times C}$ is modulated by the suppressive interactions. Depending on the hGRU unit type (feedforward or feedback), this input comes from lower or higher layers of the network. Suppression is implemented as a combination of subtractive and multiplicative influences:

$$S = \left[Z - \left((\alpha H[t-1] + \mu) \odot C^S\right)\right]_+, \quad (4)$$

where $\alpha$ and $\mu$ are learnable scalar parameters that control the strength of suppression, $[.]_+ = \max(0, .)$ denotes rectification (ReLU). The resulting state S serves as the input to the facilitation stage.

*Facilitation stage:* The intermediary state S is next used to compute facilitative interactions that enhance activity consistent with the input and its spatial context. A facilitation gate $G^F \in [0,1]^{H \times W \times C}$ is computed from the intermediate state:

$$G^F = sigmoid(U^F * S) \quad (5)$$

where $U^F \in \mathbb{R}^{1 \times 1 \times C \times C}$ denotes learnable convolutional kernels. The resulting gate $G^F$ determines, for each spatial location and feature channel, the degree to which newly computed activity is integrated into the hidden state.

The facilitation signal $C^F$ is computed as

$$C^F = GN(W^F * S) \quad (6)$$

where $W^F \in \mathbb{R}^{K \times K \times C \times C}$ denotes learnable convolutional kernels for facilitation, and *GN*(·) denotes group normalization. We used spatial kernels with *K*=5 for horizontal and *K*=1 for feedback, channel mixing interactions.

The facilitation signal $C^F$ is combined through additive and multiplicative facilitation with the intermediate state *S* to form a candidate updated state:

$$\widetilde{H} = [\nu(C^F + S) + \omega(C^F \odot S)]_+ , \quad (7)$$

where $\nu$ and $\omega$ are learnable scalars that weight additive and multiplicative facilitation, respectively. The candidate state $\widetilde{H}$ represents the newly computed activity pattern at time *t*, which is combined with the previous state *H[t−1]* based in the next time step. The candidate state is then integrated with the previous state through a gated update:

$$H[t] = (1 - G^F) \odot H[t-1] + G^F \odot \widetilde{H} . \quad (8)$$

Where $G^F$ controls the balance between retaining the previous activity *H[t−1]* and the newly computed state $\widetilde{H}$. This gating mechanism enables the hGRU to propagate and refine activity patterns over time, allowing enhanced activity to spread across object representations.

Following ref. [41], we included an additional learnable gate β for the top-down hGRUs to regulate the influence of higher-level signals. This gate controls the extent to which lower-level representations are updated by top-down recurrent interactions versus preserved from the previous state, allowing the network to determine when higher-level information should influence lower-level activity. Specifically, for top-down hGRUs only, the hidden state at layer l is updated as:

$$H^l = \left(1 - sigmoid(\beta)\right) \odot hGRU(H^l, H^{l+1}) + sigmoid(\beta) \odot H^l \quad (9)$$

where $l$ indexes the layer number. When $sigmoid(\beta)$ is small, top-down feedback strongly influences the update of $H^l$; when $sigmoid(\beta)$ is large, the lower-level representation is largely preserved. Table S1 provides the architectural details.

| Layer | Operation | Output Shape |
|---|---|---|
| **hGRU-1-horizontal** | hGRU – horizontal | $H/4 \times W/4 \times 64$ |
| **conv-1-down** | $2 \times \begin{cases} conv1 \times 1/1 \\ groupnorm \end{cases}$ | $H/4 \times W/4 \times 128$<br>$H/4 \times W/4 \times 128$ |
| | maxpool 2 x 2 / 2 | $H/8 \times W/8 \times 128$ |
| **hGRU-2-horizontal** | hGRU – horizontal | $H/8 \times W/8 \times 128$ |
| **conv-2-down** | $4 \times \begin{cases} conv1 \times 1/1 \\ groupnorm \end{cases}$ | $H/8 \times W/8 \times 256$<br>$H/8 \times W/8 \times 256$ |
| | maxpool 2 x 2 / 2 | $H/16 \times W/16 \times 256$ |
| **hGRU-3-horizontal** | hGRU – horizontal | $H/16 \times W/16 \times 256$ |
| **conv-3-down** | $4 \times \begin{cases} conv1 \times 1/1 \\ groupnorm \end{cases}$ | $H/16 \times W/16 \times 512$<br>$H/16 \times W/16 \times 512$ |
| | maxpool 2 x 2 / 2 | $H/32 \times W/32 \times 512$ |
| **hGRU-4-horizontal** | hGRU – horizontal | $H/32 \times W/32 \times 512$ |
| **conv-4-up** | bilinear interpolation | $H/16 \times W/16 \times 512$ |
| | groupnorm | $H/16 \times W/16 \times 512$ |
| | conv 1 x 1 / 1 | $H/16 \times W/16 \times 256$ |
| | conv 1 x 1 / 1 | $H/16 \times W/16 \times 256$ |
| **hGRU-5-top-down** | hGRU – top-down | $H/16 \times W/16 \times 256$ |
| **conv-5-up** | bilinear interpolation | $H/8 \times W/8 \times 256$ |
| | groupnorm | $H/8 \times W/8 \times 256$ |
| | conv 1 x 1 / 1 | $H/8 \times W/8 \times 128$ |
| | conv 1 x 1 / 1 | $H/8 \times W/8 \times 128$ |
| **hGRU-6-top-down** | hGRU – top-down | $H/8 \times W/8 \times 128$ |
| **conv-6-up** | bilinear interpolation | $H/4 \times W/4 \times 128$ |
| | groupnorm | $H/4 \times W/4 \times 128$ |
| | conv 1 x 1 / 1 | $H/4 \times W/4 \times 64$ |
| | conv 1 x 1 / 1 | $H/4 \times W/4 \times 64$ |
| **hGRU-7-top-down** | hGRU – top-down | $H/4 \times W/4 \times 64$ |
| **Readout** | conv 1 x 1 / 1 | $H/4 \times W/4 \times 128$ |
| | conv 1 x 1 / 1 | $H/4 \times W/4 \times 14$ |
| | bilinear-resize | $H \times W \times 14$ |

**Table S1. GammaNet architecture.** The network consists of a hierarchy of hGRU layers operating at progressively coarser spatial resolutions (down-sampling pathway), and corresponding upsampling layers that enable top-down interactions. "Down" denotes downsampling blocks that reduce spatial resolution via max pooling, whereas "up" denotes upsampling blocks that increase spatial resolution using bilinear interpolation. Kernel sizes are specified as height × width / stride. Horizontal hGRU layers implement recurrent interactions within a spatial scale, whereas top-down hGRU layers integrate information across scales. hGRU units use softplus nonlinearities, and convolutional layers use ReLU activations.

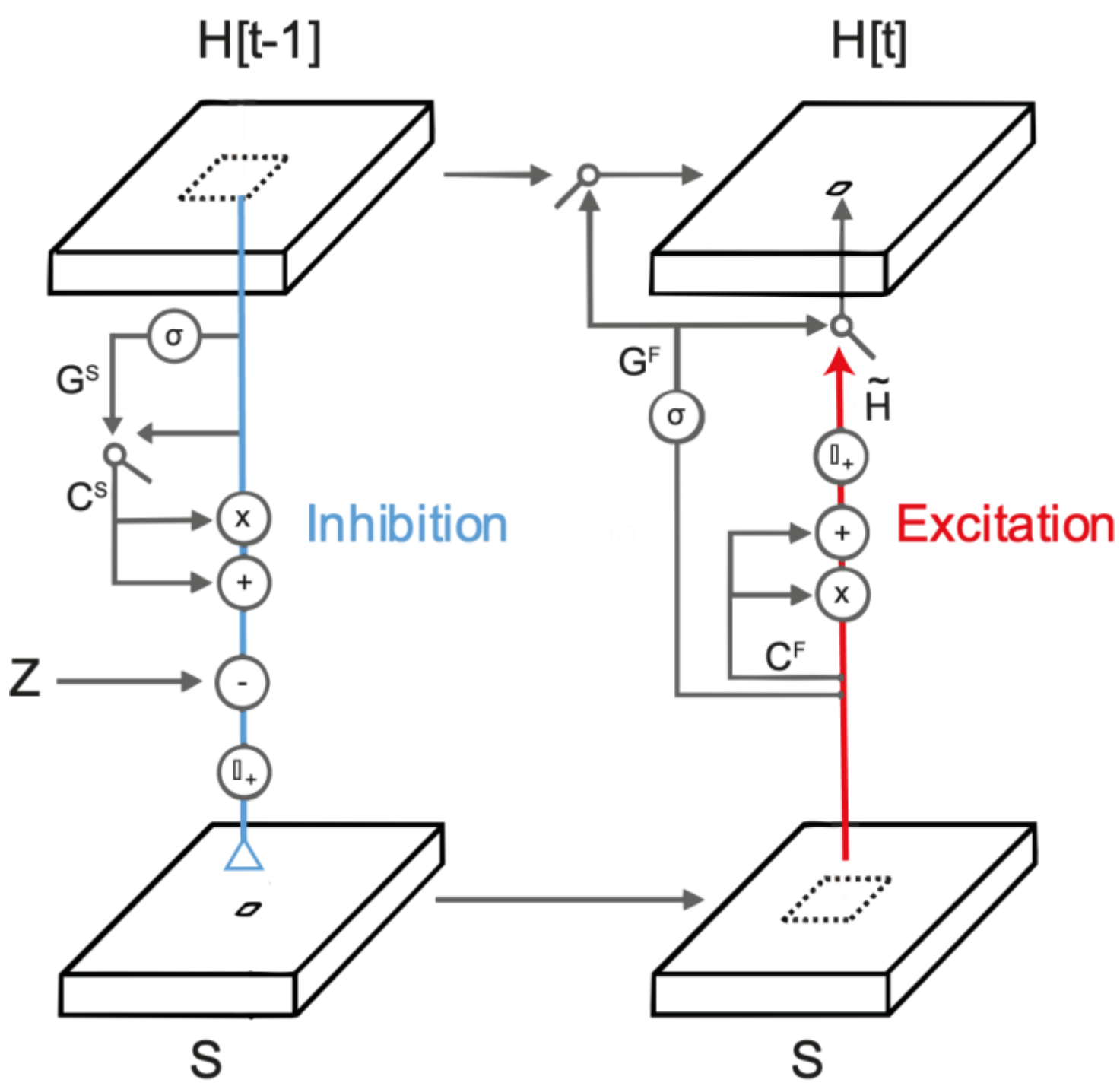

**Figure S1. Architecture of the hGRU unit**. Schematic of the hGRU update at a single timestep, illustrating the suppression (left, blue) and facilitation (right, red) stages. The previous hidden state $H[t-1]$ interacts with the current input $Z$ through a sequence of gated operations. In the suppression stage, the gate $G^S$ modulates the influence of the previous state to compute a suppressive signal $C^S$, which reduces the input to form the intermediate state $S$. In the facilitation stage, the intermediate state $S$ drives excitatory interactions via $C^F$, generating a candidate state $\widetilde{H}$. The facilitation gate $G^F$ then determines how the candidate state is integrated with the previous state to produce the updated hidden state $H[t]$. Details of the interactions are described in the Methods section describing the hGRUs. Adapted from Linsley et al., 2018[30].

## C-RBP

We trained the network using contractor recurrent backpropagation (C-RBP)[33] an extension of recurrent backpropagation (RBP)[34–36]. In standard RBP, the recurrent network is first run following

$$h_{t+1} = F(x, w_F, h_t) \tag{10}$$

Where $h_t$ is the hidden state at time t, $x$ is the input, and F denotes the recurrent network parametrized by $w_F$. The network is run until its activity converges to a fixed point. At equilibrium,

$$h^* = F(x, w_F, h^*) \tag{11}$$

Rather than unrolling the network through time and backpropagating through every step, as in BPTT, RBP introduces a second linear accessory network with an extra unit $z$ for every unit in $h$ whose connectivity is the transpose of the original network's connectivity. This accessory network propagates credit assignment signals and receives the loss gradient $\frac{\partial L}{\partial h^*}$ as input. Denoting this accessory network's state by z,

$$z_{t+1} = \mathrm{J}^{\top}.z_t + \frac{\partial L}{\partial h^*} \tag{12}$$

Where J is the Jacobian matrix of F evaluated at $h^*$; see ref.[37] for a biologically plausible implementation. At equilibrium, this second network converges to:

$$z^* = \mathrm{J}^{\top}.z^* + \frac{\partial L}{\partial h^*} \tag{13}$$

Since both $h^*$ and $z^*$ are fixed points, gradient computation depends only on these converged values and not on the intermediate states that BPTT must retain, making RBP local in time.

Moreover, the gradient with respect to the network's parameters takes the form

$$\frac{\partial L}{\partial w_F} = z^{*\top} \frac{\partial F(x, w_F, h^*)}{\partial w_F} \tag{14}$$

requiring only the equilibrium $z^*$ and the local dependence of F on its own parameters, evaluated at $h^*$, no propagation of error terms across the network is needed, making RBP local in space.
For the case $F(x, w_F, h^*) = f(Wh + Ux + b)$ where f is applied elementwise and $u^* = Wh^* + Ux + b$ is the unit's input at equilibrium, this reduces to:

$$\frac{\partial L}{\partial W_{i,j}} = z_i^* f'(u_i^*) h_j^* \tag{15}$$

Each synaptic update is the product of the presynaptic activity $h_j^*$ and the postsynaptic error $z_i^* f'(u_i^*)$, both quantities available locally at the synapse[37].

To ensure that the activity of the recurrent neural network converges to a fixed point, enabling computation of gradients with respect to this steady state, we used a variant of RBP, C-RBP, which adds a Lipschitz Coefficient Penalty (LCP) to the loss term[33]. This regularization (weight $10^{-4}$) encourages the recurrent update function to be locally contractive around its fixed point, ensuring stable dynamics while preserving expressive transient dynamics.

Formally, let $f(\cdot; W)$ denote the recurrent update function with parameters $W$, and let $\widehat{H}$ be a fixed point such that $\widehat{H} = f(\widehat{H}; W)$. Stability requires that the Jacobian of the update function with respect to the hidden state, $J_{f,\widehat{H}}$, has a spectral radius less than one. Directly computing or constraining this Jacobian is infeasible for high-dimensional recurrent networks, as its size grows quadratically with the dimensionality of the state. The LCP circumvents this limitation by constraining an efficient approximation to the Jacobian norm using a vector–Jacobian product. Specifically, it penalizes large column-wise sums of the Jacobian by minimizing

$$\mathcal{L}_{\mathrm{LCP}}(W) = \| \left(\mathbf{1} \cdot J_{f,\widehat{H}} - \lambda\right)_+ \|_2^2, \tag{16}$$

where $\mathbf{1}$ is a vector of ones and $\lambda \in [0,1)$ controls the degree of contraction. This formulation provides an efficient upper bound on the Jacobian norm and can be computed using automatic differentiation without explicitly forming the Jacobian matrix[33]. The LCP term was added to the task loss during training. Importantly, enforcing local contraction at the fixed point does not eliminate transient dynamics. The network can still exhibit rich, time-varying activity before convergence, including the gradual spread of enhanced activity underlying perceptual grouping.

## Training of the GammaNet

We trained the GammaNet using the Adam optimizer with a batch size of 28 images distributed across four NVIDIA RTX GPUs (24 GB memory each). Input images were resized to 224×224, and training proceeded for 21,000 iterations using a pixel-wise cross-entropy loss (Eq. 1). The learning rate followed a piecewise constant schedule with linear warmup: starting from zero, the learning rate increased linearly to $10^{-4}$ over the first 1,000 iterations, then remained constant until iteration 10,000 and 20,000, at which point it was reduced by a factor of 0.8, yielding learning rates of $8.10^{-5}$ and finally $6.4.10^{-5}$. The ResNet-101 FPN backbone, pre-trained on COCO, was kept frozen throughout training, with gradients flowing only through the recurrent head and segmentation decoder. The recurrent head operated for 40 timesteps. During weight update, the Jacobian penalty regularizer was applied as described in the previous paragraph[33]. Model performance was evaluated on the validation set every 2,000 iterations. We performed five independent training runs and report results of all models.

## Image set

We used a custom image set derived from the COCO image segmentation dataset[40], following the stimulus design of ref. [22]. Although this dataset includes 41 object classes, their relative frequencies are imbalanced[22,40]. Several classes are underrepresented (e.g. “fire hydrant”) and we found that the network performance degraded for these rare categories. To ensure reliable training and evaluation, we restricted the image set to the 13 most well-represented classes, which together form 84% of the dataset of ref. [22]. Conversely, the class “person” was strongly overrepresented (Figure S2). To reduce this imbalance, we randomly subsampled one sixth of the available “person” instances. For each selected object covering at least 5000 pixels, we placed cues at three randomly chosen locations within the object. When an image contained multiple instances from the selected classes satisfying the size criterion, it was included multiple times, once for each valid object-cue combination.

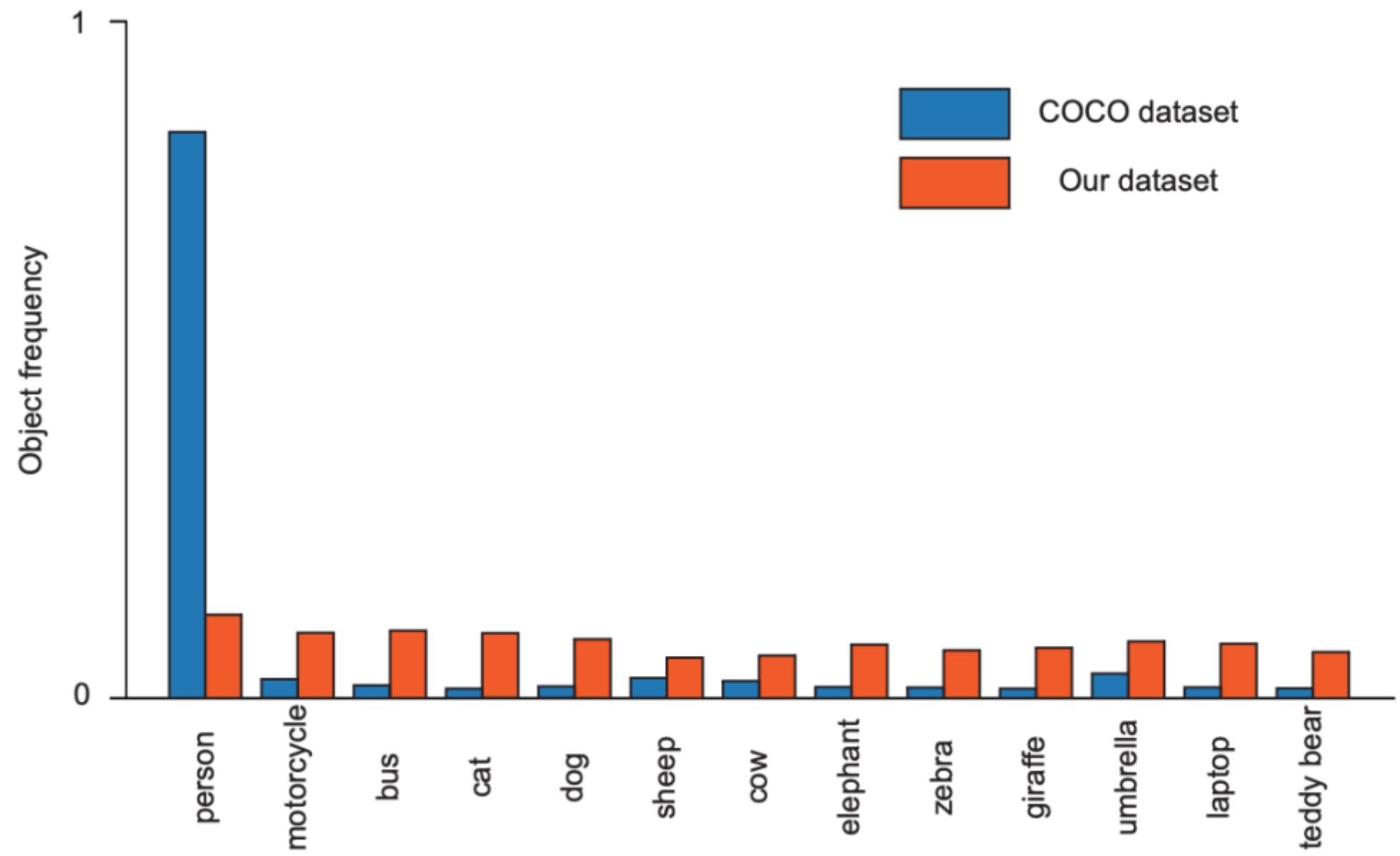

**Figure S2.** Frequency of classes in the original COCO dataset (blue) and in our dataset (orange).

## Evaluation of the GammaNet on the image set

The images used by Adeli et al. [22] to measure human RTs were drawn from the test set of the COCO segmentation dataset and we did not include them in the image set used to train the GammaNet. Adeli et al. [22] added a fixation point and a cue at a distance of either 3 or 6 d.v.a. from the fixation point. In our simulations, we used the same images but presented only the fixation point.

For each image-cue combination, we considered the GammaNet’s output correct if both of the following criteria were met:

- The IoU between the predicted and ground truth segmentation masks exceeded 0.5.
- The image regions corresponding to the cue and the target were classified correctly and included in the predicted segmentation mask

We chose IoU over panoptic quality (PQ), the metric more commonly used in general-purpose panoptic segmentation benchmarks[62], because with a single ground-truth class per image, PQ reduces to a thresholded version of IoU: it equals IoU when IoU > 0.5, and zero otherwise. This measure discards meaningful information below the threshold and introduces an unnecessary discontinuity.

Under these criteria, the network achieved an accuracy of 78% on the same-different task, compared to a human accuracy of 90%. The difference partly reflects the more conservative evaluation of the model, because we required correct segmentation and classification of all relevant regions, whereas human observers only needed to judge whether the cue fell on the same object as the fixation point. Future work could directly predict same vs. different, providing a more comparable accuracy measure.

## Adversarial hidden state generation

To generate adversarial hidden states, we used the Basic Iterative Method (BIM)[47], which is an iterative extension of the Fast Gradient Sign method (FGSM)[64]. In contrast to standard adversarial attacks on images, we directly perturbed the activity of units in the highest hGRU layer so that a linear classifier misclassified the representation:

$$X^{adv} = X + \varepsilon \cdot sign\left(\nabla_X J(X, Y_{true})\right) \tag{17}$$

where $X^{adv}$ denotes the adversarial hidden state, $X$ the original hidden state, $\varepsilon$ is the perturbation magnitude and $\nabla_X J(X, Y_{true})$ the gradient of the loss function with respect to the hidden state. We define $J(.)$ as the cross-entropy loss between the predicted class of the linear probe and the ground-truth label. The BIM applies this update iteratively using a small step size $\alpha$, yielding a sequence of adversarial hidden states:

$$X_0^{adv} = X \tag{18}$$

$$X_{N+1}^{adv} = Clip_\varepsilon\left(X_N^{adv} + \alpha \cdot sign\left(\nabla_X J\left(X_N^{adv}, Y_{true}\right)\right)\right) \tag{19}$$

where $Clip_\varepsilon(\cdot)$ constrains the perturbed hidden state to lie within an $\varepsilon$-ball around the previous perturbed hidden state, i.e. $[X - \varepsilon, X + \varepsilon]$. We used N = 5, $\alpha = 0.025$, $\varepsilon = 0.25$.

## Timing of the semantic feedback signal for internal edge crossing

To determine when top-down signals are required to cross internal edges (Fig. 4,5), we performed temporally selective perturbations of the top hGRU layer. The perturbation (BIM) that degrades semantic information was applied just before the segmentation mask reached the internal edge and maintained for five timesteps, after which it was removed. During this period, the segmentation mask propagated up to the internal edge but stalled there. Once the perturbation was lifted and semantic information was restored, propagation resumed and the grouping process crossed the edge.

To quantify this effect across the example edges, we defined propagation speed as the increase in the fraction of the after-edge region covered per timestep. Across images, propagation consistently slowed and often halted during perturbation (Fig. S4). Once the perturbation was removed, propagation resumed. These results demonstrate that top-down semantic signals are required specifically when internal boundaries are crossed.

## Estimation of human reaction times

To relate model dynamics to human RTs, we presented images with the fixation cue, and tracked the propagation of the segmentation signal over time. We defined the model RT as the first timestep at which the location of the second cue was correctly labeled, and required that it remained correctly labeled at the end of the propagation process. RTs were analyzed only for image-cue combinations for which the GammaNet produced a correct segmentation (78%, see previous section).

Human RTs were averaged across participants for each image–cue condition (255 images × 2 cue positions), following Adeli et al.[22]. To quantify the correspondence between model and human RTs, we performed a linear regression and computed the coefficient of determination ($R^2$) for each model. Differences in explained variance between models were assessed using standard errors of $R^2$ estimates and corresponding Z-tests[65].

To estimate the maximum explainable variance given the inter-participant variability we computed the $R^2$ of a linear regression predicting one group's averaged RTs from the other's. We repeated this procedure 50 times, and the mean $R^2$ across splits is reported as the noise ceiling in Table 1.

## Image credits

The natural images shown in the figures are taken from the Microsoft COCO dataset[40] and were originally published on Flickr by their authors under Creative Commons licenses. In all figures, the images were modified by adding a fixation cue and by overlaying model outputs (predictions, activity or masks). The original photographs remain under their respective licences listed below; the modified versions shown here are distributed under the licence of this article (CC BY-NC-SA 4.0).

**Figs. 1H, 2 and 3:** “At our favorite pizza place in Moraira” by MollySVH (COCO image 286553; https://www.flickr.com/photos/18472724@N00/4252543190/), licensed under CC BY 2.0 (https://creativecommons.org/licenses/by/2.0/); modified (fixation cue added, model outputs overlaid).

**Figs. 4 and 5:** “The Alliance Sessions #12” by Andrea Marutti (COCO image 560911; https://www.flickr.com/photos/afeman/2157319947/), licensed under CC BY-NC-SA 2.0 (https://creativecommons.org/licenses/by-nc-sa/2.0/); modified (fixation cue added, model outputs overlaid).

**Fig. 4A:** “Ayutthaya Elephants” by Dustin Quasar (COCO image 170739; https://www.flickr.com/photos/dustinq/7244940442/), licensed under CC BY-NC-SA 2.0 (https://creativecommons.org/licenses/by-nc-sa/2.0/); modified (fixation cue added, model outputs overlaid).

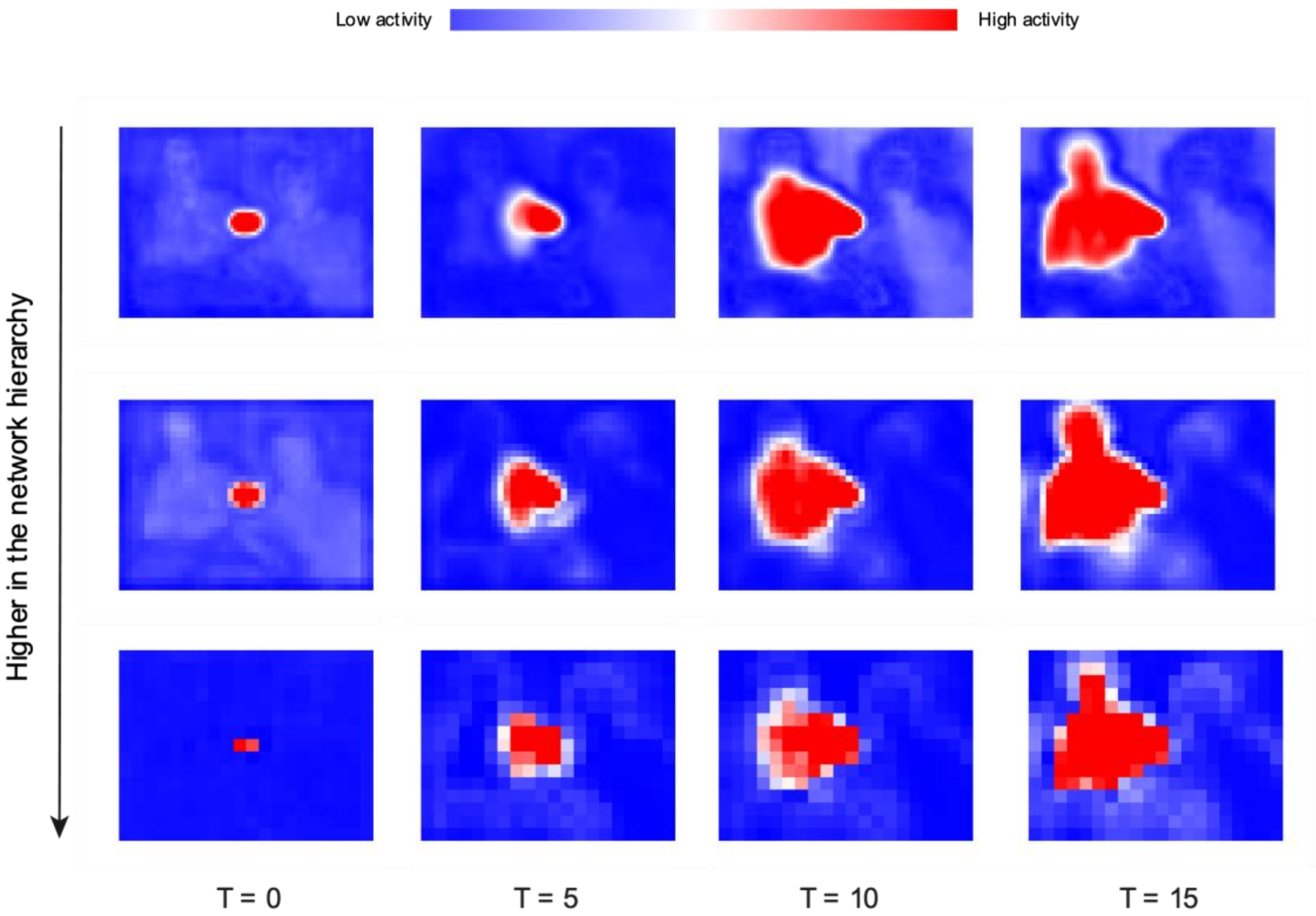

**Figure S3.** Activity of hGRUs across the network hierarchy for an example image. Enhanced activity spreads across the representation of the target object, while activity outside the object remains low.

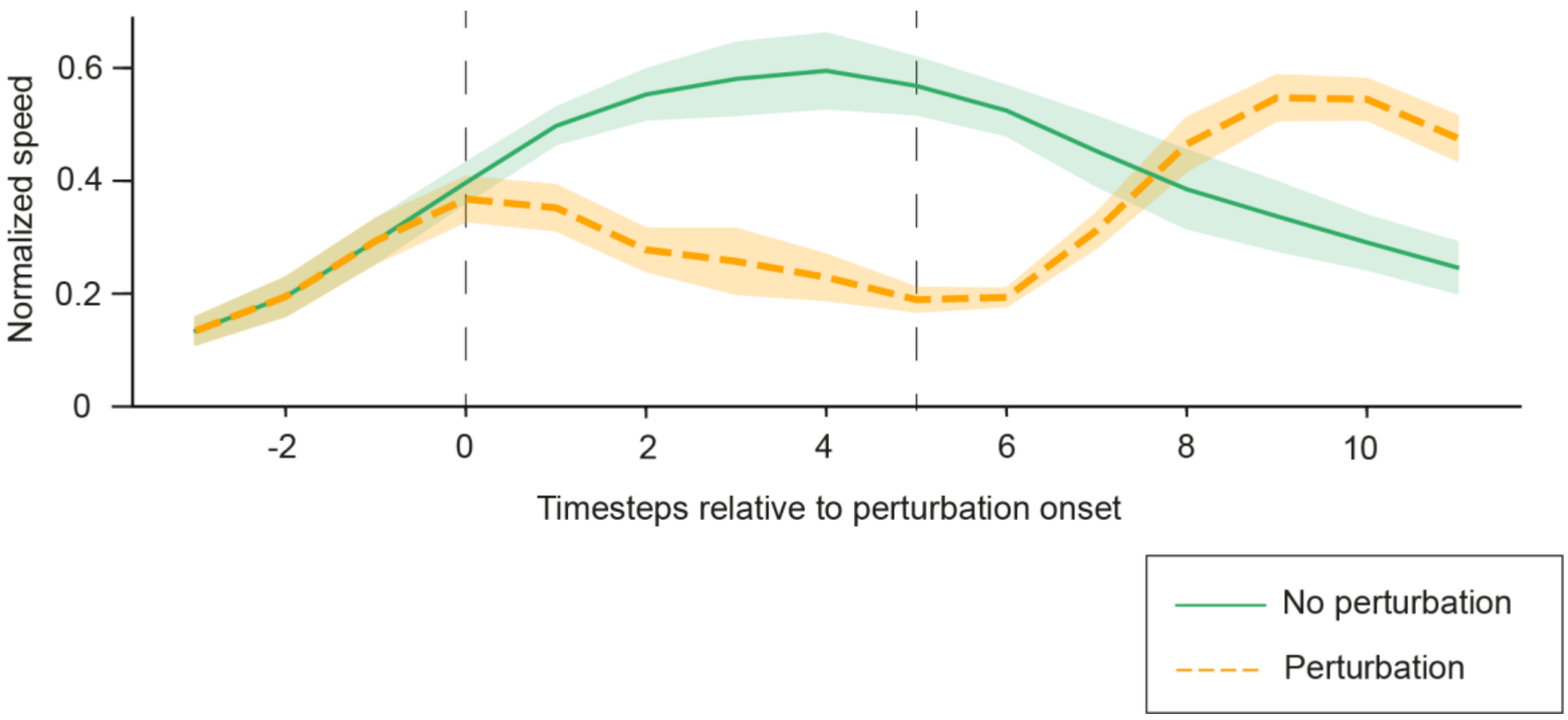

**Figure S4.** Propagation speed of the grouping signal into the after-edge region, with (orange traces) and without perturbation (green traces). For each image, perturbation onset (t=0) occurred at the timestep at which the grouping signal reached the internal edge, and the perturbation was maintained for five timesteps (dashed lines). Shaded bands around each trace show ± SEM. The perturbation causes the propagation speed to become almost zero and the speed recovers once the perturbation is removed